\documentclass{iau}
\usepackage{graphicx}
\usepackage{natbib}

\def\ga{\mathrel{\mathchoice {\vcenter{\offinterlineskip\halign{\hfil
$\displaystyle##$\hfil\cr>\cr\sim\cr}}}
{\vcenter{\offinterlineskip\halign{\hfil$\textstyle##$\hfil\cr
>\cr\sim\cr}}}
{\vcenter{\offinterlineskip\halign{\hfil$\scriptstyle##$\hfil\cr
>\cr\sim\cr}}}
{\vcenter{\offinterlineskip\halign{\hfil$\scriptscriptstyle##$\hfil\cr
>\cr\sim\cr}}}}}

\def\apj{ ApJ}
\def\aap{ A\&A}
\def\mnras{MNRAS}

\def\araa{ARAA}
\def\aj{AJ}
\def\apjs{ApJ Supp}
\def\apjl{ ApJL}

\def\aapr{Astronomy and Astrophysics Review}

\title[Evolution of binaries with compact objects in GCs] 
{Evolution of binaries with compact objects in globular clusters}

\author[Natalia Ivanova]   
{Natalia Ivanova$^1$}

\affiliation{$^1$ Dept.\ of Physics, University of Alberta,  11322-89 Ave, Edmonton, AB, T6G 2E7, Canada \\ email: {\tt nata.ivanova@ualberta.ca}}

\pubyear{2014}
\volume{312}  
\pagerange{203--212}
\jname{Star clusters and black holes in galaxies across cosmic time}
\editors{Y. Meiron, S. Li, F.-K. Liu \& R. Spurzem, eds.}
\doi{10.1017/S1743921315007826}

\begin{document}

\maketitle

\begin{abstract}
Dynamical interactions that take place between objects in dense
stellar systems lead to frequent formation of exotic stellar objects,
unusual binaries, and systems of higher multiplicity. They are most
important for the formation of binaries with neutron stars and black
holes, which are usually observationally revealed in mass-transferring binaries.  
Here we review the current understanding of
compact object's retention, of the metallicity dependence on the formation of low-mass
X-ray binaries with neutron stars, and how mass-transferring binaries with a black hole and a white dwarf 
can be formed.  We discuss as well one old unsolved puzzle and two  
new puzzles posed by recent observations: what descendants do ultra-compact X-ray binaries
produce,  how are very compact triples formed,   and how can black hole low-mass X-ray binaries 
acquire  non-degenerate companions?

\end{abstract}

\firstsection 
\section{Introduction}

A low-mass X-ray binary (LMXB) is a binary star consisting of a
neutron star (NS) or a black hole (BH) accretor, with a less-massive
donor that fills its Roche lobe (RL).  A donor can be a main-sequence
(MS) star, a white dwarf (WD) or a red giant (RG).  An isolated binary
(a field binary) can become an LMXB via several possible scenarios
\citep[e.g.,][]{1991PhR...203....1B,1993ARA&A..31...93V,2006book}.  A
typical formation scenario starts with a common envelope event, where
the initially most massive star overfills its RL.  The expulsion of
the formed common envelope leads to the binary orbit
tightening. Later, the initially most massive star explodes in a
core-collapse (CC) Ib or Ic supernova and the compact object is
formed. Further orbit's shrinkage can take place due to tides,
magnetic braking and gravitational waves in, likely, an eccentric
binary; a second episode of a common envelope event can take place as
well. When the initially less massive stars overfills its RL, the
system is revealed as an X-ray source.  It is essential that a typical
formation scenario of a field LMXB includes several phases or
processes which are not yet fully understood -- tides, common envelope
events, magnetic braking and natal supernova kicks (see, e.g., reviews
\citealt{2006epbm.book.....E,2006book,2008EAS....29...67Z,
  2011ApJS..194...28K, 2013A&ARv..21...59I}). We further distinguish
LMXBs by how they appear -- whether they have a stable disk accretion
and therefore can be seen persistently, or if they raise drastically
their X-ray luminosity in outbursts, and therefore are transient
sources.

Globular clusters (GCs) contain about two orders of magnitude more
LMXBs per stellar mass than the Milky Way (MW) population contains of
``native'' LMXBs
\citep{1984AdSpR...3...19G,1995book,2004ApJ...607L.119B}.  This
overabundance has been attributed to the role that dynamical
encounters play in the formation of LMXBs in GCs.  Indeed, studies
have shown strong positive correlation between the stellar encounter
rate and the number of close X-ray binaries in GCs
\citep{2003Pooley,2013ApJ...766..136B}.  Additionally, the {\it
  metallicity dependence} was found -- that a metal-rich GC in the
Galaxy and in M31 is $\sim 3$ times more likely to contain a bright
LMXB than a metal-poor GC
\citep{1993ASPC...48..156G,1995ApJ...439..687B}.  Observations of GCs
around distant galaxies have also shown that the fraction of GCs
hosting LMXBs is larger by a factor of ~3 in metal-rich than in
metal-poor GCs
\citep{2002ApJ...574L...5K,2003ApJ...595..743S,2006ApJ...647..276K,
  2007ApJ...660.1246S,2011ApJ...736...90P,2013ApJ...764...98K}.

It is interesting that most of the very uncertain physics involved in
the formation of ``native'' field LMXBs -- e.g., common envelope
phase, binary modification due to natal supernova kicks, tides -- is
either not involved in the formation scenario of a dynamically formed
LMXB, or affects the formation scenario with a different, often
lesser, impact.  For example, supernova kicks affect GCs LMXBs mainly
by reducing the retention of compact objects in GCs after their
formation \citep{1987IAUS..125..187V,1996MNRAS.280..498D}.  The
formation path of an LMXB in a GC, on the other hand, can take a
short-cut by pairing a NS or a BH via a dynamical encounter. To
understand the puzzles that observations bring us, we need to analyze
how the dynamical formation channels -- binary exchanges, tidal
captures, physical collisions and triple formations -- perform with
different type of donors.

\section{Formation and retention of compact objects}

The mean three-dimensional pulsar birth velocity is 400 km/s, and a
pulsar  two-dimensional speed can be as large as 1600 km/s
\citep{2005MNRAS.360..974H}.  The escape velocity from a GC is an
order of magnitude less than pulsars' mean kick velocities. If all NSs
are born with kicks we derived for pulsars, only a tiny fraction of
them could remain in a GC \citep{2002ApJ...573..283P}.  Indeed, a
typical GC with current mass of $2\times10^5 M_\odot$ and Kroupa's
initial mass function \citep{2002Sci...295...82K}, forms about 3000
NSs. If the escape velocity is 40km/s, only one NS can be retained if
all stars were single at birth \citep{2008MNRAS.386..553I}. Being born
in a binary changes the post-kick system's space velocity
\citep[e.g.,][]{1995MNRAS.274..461B,2002ApJ...573..283P}, and if all
stars would be initially in binaries, a $2\times10^5 M_\odot$ GC could
retain 15 NSs \citep{2008MNRAS.386..553I}. This is still too little to
explain the observations.  Indeed, Ter~5, which is about $2\times10^6
M_\odot$, is estimated to contain 150 recycled pulsars
\citep{2011MNRAS.418..477B}.

The problem is considered now to be resolved by assuming that the
observed high pulsar velocities are only related to the case of a CC
supernova.  Indeed, there is another formation channel for NSs,
electron capture (EC) supernova, that takes places when a degenerate
ONeMg core reaches $1.38M_\odot$
\citep{1980PASJ...32..303M,1984ApJ...277..791N,1992ApJ...396..649T}.
EC supernova takes place in a narrow range of masses in single stars,
where this range depends slightly on metallicity and the used stellar
code (for example, this range is $7.7-8.3M_\odot$ at $Z=0.02$ and
$6.2-6.8 M_\odot$ at Z=0.0005 using the STARS/ev code,
\citealt{1995MNRAS.274..964P}).  In binary stars, the range of initial
stellar masses that can produce EC supernova is larger, due to mass
transfer history \citep{2004ApJ...612.1044P}.  In addition, for stars
in binaries, EC NSs can be formed due to accretion on a companion WDs,
or due to a merger of two WDs.  As a result, in field, 10-15\% of all
NSs could be formed via various EC channels
\citep{2008MNRAS.386..553I}.  Probably the most famous example is Crab
Supernova \citep{2006A&A...450..345K,2013ApJ...771L..12T}.  The most
important feature of EC supernovae for GCs is that the explosion is
much weaker \citep[e.g.,][]{2006ApJ...644.1063D}, and hence their
kicks are believed to be weaker as well. Under assumption of lower
kicks, most of retained NSs in a GC come from different EC supernovae
channels. The ratio of CC NSs to EC NSs in a GC is about 1 to 30-200
instead of about 10 to 1 in the field \citep{2008MNRAS.386..553I}.
Since post-EC NS masses are low, $1.22-1.27 M_\odot$, a low-mass
dominated NSs mass function can be expected in GCs.

The understanding of BHs retention has recently drastically changed.
It can be estimated that each $150-200~M\odot$ of the current, “aged”,
stellar mass, have produced a BH in the past.  This is assuming
Kroupa's initial mass function and no mass loss from a GC except that
due to stellar evolution.  Due to lower kicks, the retention fraction
immediately after formation is much higher than for NSs, 30-40\%, if
the escape velocity from a GC is 50 km/s \citep{2006ApJ...650..303B}.
It was considered previously that equipartition between the stars in a
GC would lead to the formation of a BH sub-cluster in a center of a GC
(Spitzer instability), followed by rapid BH population depletion via
an ejection of nearly all BHs
\citep{1969ApJ...158L.139S,1993Natur.364..421K}.  However, recent
detailed numerical simulations of BH subclusters have shown that up to
~20\% of the BHs may remain in massive BH subclusters, and these
subclusters do not reach equipartition
\citep{2006ApJ...637..937O}. Monte Carlo simulations of a whole GC
have shown that up to 25\% of initial BHs can remain and participate
in interactions with other stars \citep{2010MNRAS.407.1946D}, or that
even more than a half of formed BHs can remain, with no evidence for
the Spitzer instability \citep{2013ApJ...763L..15M}.  It can be now
assumed that a large number of initially retained BHs can remain in
GCs, 10-60\%, depending on cluster mass and its initial virial radius
\citep{2014arXiv1409.0866M}.

\section{Formation Channels and Metallicity dependence}

The observed metallicity dependence for bright LMXBs can be
interpreted as a combination of the metallicity dependence due to the
encounter rates, and the metallicity dependence due to the mass
transfer -- whether the formed LMXBs appear as persistent or transient
sources, how bright they are when 'on', and how long they can remain as bright sources.

Typical dynamical formation channels of a binary with a compact object
are: (i) via binary exchanges (BEs), (ii) physical collisions (PCs)
with a giant or (iii) via tidal captures (TCs) \citep[e.g., see ][and
  references therein]{2006csxs.book..341V}.  A dynamical event provides
a {\it direct} LMXB formation if a newly formed binary can start the
mass transfer within the Hubble time while evolved in isolation.  Not
all formed binaries with a NS and a BH will become LMXBs directly, but
the formed binaries can serve as seed binaries for further dynamical
encounters (hardening, triples formation) and eventually can also
start the mass transfer.

For MW~GCs, bright LMXBs are those that have their X-ray luminosity
$L_X > 10^{36}$~erg~s$^{-1}$.  18 bright transient or persistent LMXBs have been observed in MW~GCs. For the 13
known by 2004, a compilation is provided in
\citealt{2006csxs.book..341V}.  Two more transients were found in Ter~5
\citep{2010ATel.2919....1B,2014ApJ...780..127B}, one more in NGC~6440
\citep{2010ApJ...714..894H}, and the first transient was found in each of two GCs, 
NGC~6388 and M~28
\citep{2011A&A...535L...1B,2013ATel.4925....1E,2013Natur.501..517P}. 

The separation between metal-rich (also known as red) and metal-poor
(also known as blue) MW~GCs is not strictly defined, and is usually
taken at [Fe/H]=-1 (a gap in the distribution of metallicities is observed between [Fe/H]=-1.2 and [Fe/H]=-0.8, and
this gap disappears if the sample is limited to low-reddening clusters,
\citealt{2014MNRAS.437.1725V,2014MNRAS.437.1734V}).  Here, we will
refer to the GCs that have [Fe/H] between -1.5 and -0.5 as 
``intermediate'' metallicity GCs, if [Fe/H]$<-1.5$ as  ``distinctly
metal-poor'' GCs, and if [Fe/H]$>-0.5$ as  ``distinctly metal-rich''
GCs.

From 18 known bright MW~GCs LMXBs only 2 are in ``distinctly
metal-poor'' GCs (both are in M~15), 8 are in the ``intermediate''
metallicity GCs (LMXBs in M~28, NGC~1851, NGC~6388, NGC~6652,
NGC~6712, Ter~1, Ter~2, Ter~6) and another 8 are in ``distinctly
metal-rich'' GCs (LMXBs in Ter~5, NGC~6440, NGC~6441, NGC~6624,
Lil~1), where two GCs contain more than one bright LMXB (Ter~5 and
NGC~6440).

If we look at ``classes'' of bright MW~LMXBs by their donor-type (note
that this is also strongly linked to their orbital periods), we can
see that 6 likely have a WD donor \citep{2006csxs.book..341V,2009ApJ...699.1113Z, 2010ApJ...712L..58A}. These binaries are called
ultra-compact X-ray binaries (UCXBs).  3 of these UCXBs are in
``intermediate'' GCs (NGC~1851, NGC~6712, Ter~2), one in a
``distinctly metal-poor'' GC (M~15) and two in ``distinctly
metal-rich'' GCs (NGC~6624, NGC~6440).  2 of bright MW~LMXBs likely
have MS donors, both are persistent and are located in GCs with
[Fe/H]$>-1$ (NGC~6441, \citealt{2004MNRAS.347..334B}, NGC~6652, 
\citealt{2012ApJ...747..119E}).  5 of bright LMXBs have most likely
either a subgiant or a giant companion \citep{2006csxs.book..341V, 2012A&A...547A..28T, 2013Natur.501..517P}, one in a ``distinctly
metal-poor'' GC (M~15), two in ``distinctly metal-rich'' GCs (Ter~5,
NGC~6440) and two are in the ``intermediate'' GCs (M~28, Ter~6).  In
addition, 5 bright LMXBs are transients where 
neither the companion nor orbital period has yet been determined, 
all of which are located in
metal-rich GCs with [Fe/H]$\ga-1$.  Among LMXBs where we have some knowledge of the donor, most are persistent (8 persistent versus 5 transient LMXBs).
Transient sources tend to have a longer orbital period, both for
non-degenerate donors (4 out of 5 LMXBs with giant donors are transient,
while all LMXBs with MS donors are persistent) and degenerate donors
(the UCXB with the largest orbital period, in NGC~6440, is transient).
 
As we can see, for MW~GCs, the 3 times greater probability of
hosting an LMXB by a red GC than by a blue GC is {\it cumulative}.
The ratio would change depending on what is defined as a metal-poor and
a metal-rich GC, and is not satisfied for any class of donors
separately.  The most striking difference is for LMXBs with MS donors,
but this is based on two LMXBs only.

Studies of extragalactic GCs have shown a similar preference for LMXBs
to reside in metal-rich GCs.  The ratio was found to hold across the
range of X-ray luminosities, from $2\times10^{37}$ erg s$^{-1}$ to
$5\times10^{38}$ erg s$^{-1}$ \citep{2013ApJ...764...98K}.  For even
higher X-ray luminosity, where the accretors are likely BHs, this
ratio is less certain, but still is above one, $2.5^{+0.9}_{-1.1}$.
There, it also was shown that the metallicity effect is not affected
by other factors such as stellar age, GC mass, stellar encounter rate,
and galacto-centric distance.  The caveat in making a link between
MW~GCs LMXBs and extragalactic ones is that the the boundary between
blue and red GCs is not the same, and for extragalactic GCs is done
using colors \citep[e.g.,][]{2006ARA&A..44..193B}, not [Fe/H] as for
MW~GCs. The relation between colors and [Fe/H] is however non-linear
\citep{2014MNRAS.437.1734V}.

It is important also that the popultion of detected extragalactic
LMXBs is not similar to the population of bright GC LMXBs in our Milky
Way.  Indeed, in the Galaxy, the brightest observed GC UCXB 4U1820-303,
 in NGC 6624, has $L_x \approx 4-7\times 10^{37}$~erg~s$^{-1}$
\citep{1974ApJS...27...37G,1993A&A...279L..21V} (we note that this
LMXBs could be special in other respects too, see \S\ref{triples}).
In extragalactic GC LMXBs, statistically significant studies were done
for LMXBs above $L_x \approx 2\times 10^{37}$~erg~s$^{-1}$
\citep{2013ApJ...764...98K}.  Different ranges of X-ray luminosity are
dominated by different donors \citep[e.g.,][]{2008ApJ...683..346F}.
For example, mass transfer in binaries with MS donors is driven by
magnetic braking, and rarely produce X-ray luminosity $\ga 2 \times 
10^{37}$~erg~s$^{-1}$
\citep{2008ApJ...683..346F,2009ApJ...702L.143F}.  Hence, MS donors
most likely are not present among the X-ray sources that form X-ray
luminosity functions of extragalactic GCs.

Among LMXBs, the evolution of UCXBs is best understood; predominantly
they will be seen as bright persistent sources
\citep[e.g.,][]{2003ApJ...598.1217D,2004ApJ...607L.119B}, although
there are some discrepancies between theoretically expected and
observationally derived mass-transfer rates for UCXBs with longer
orbital periods
\citep{2012A&A...543A.121V,2013ApJ...768..183C,2013ApJ...768..184H}.

The strongest uncertainty for extragalactic GC LMXBs is coming from
our limited knowledge of the duty cycles in LMXBs with giant
donors, and their luminosity during outburst \citep[see overview
  of the problem in][]{2008ApJ...683..346F}.  Further, their outburst
duration can be longer than the history of X-ray astronomy, and
hence observationally they would be known as persistent. There are
several X-ray binaries in the galaxy which confirm this suspicion as
their outburst duration is at least 20 years -- for example, GRS
1915+105 which turned on 1992 and has been active ever since
\citep{1992IAUC.5590....2C,2009MNRAS.400.1337D}, or the X-ray binary in
Ter~1, which was active for about 20 years and then turned off between 1996 and 1999
\citep{1999A&A...349..819G}.

LMXBs with RG donors are most likely to be formed via BE encounters;
the evolutionary expansion of a giant brings the system to the start
of a mass transfer.  The BEs formation channel does not have a
metallicity dependence for the encounter rates (for GCs which
otherwise are ``dynamical twins'').
The mass transfer slightly favors the appearance of metal-rich LMXBs --
those can have the short duration of a persistent mass transfer, while
metal-poor LMXBs with RG donors are likely always in quiescence
\citep{2012ApJ...760L..24I}.  It is interesting that the observations
of MW~GCs show the opposite situation to theoretical expectations: the
only persistent RG-LMXB is located in a metal-poor cluster M~15,
although this can be related to the uncertainty in our understanding
of the duty cycles.

For LMXBs with MS donors, the formation mechanisms include BEs and
TCs.  TCs, unlike BEs, can provide direct LMXB formation in most of
the formed binaries, though cumulatively are less efficient for LMXBs
formation than BEs \citep{2008MNRAS.386..553I}.  In close NS-MS
binaries, the most effective mechanism of orbital shrinkage is the
magnetic braking, which operates only in stars with an outer
convective zone.  TCs are also more effective with stars that are
convective \citep{1993A&A...280..174P}.  However, most-massive MS
stars in metal-poor GCs lack surface convective zone
\citep{2006ApJ...636..979I}.  This slight difference between stars
which are convective at the surface, and those which are radiative,
for a small mass-range, affects the following: (i) the rate of the
orbital shrinkage, where MS-NS binaries with a larger post-BE or
post-TC orbital period can start the MT if their MS star is
metal-rich; (ii) the formation rate via TCs, which are more efficient
in metal-rich GCs; and (iii) metal-poor MS-LMXBs are expected to be in
quiescence, where persistent MS-LMXBs are expected to be observed only
in metal-rich globular clusters.  The dynamically formed NS-MS
binaries that would have started the mass transfer if the donor is
metal-rich and have failed to start the mass-transfer with a
metal-poor donor, will start the MT when the donor leaves the MS.

The best-understood formation mechanism of GC~LMXBs is for UCXBs --
via PC of a NS with a RG (as was initially proposed
in \citealt{1987ApJ...312L..23V}, and see most recent details in
\citealt{2005ApJ...621L.109I,2006ApJ...640..441L}).  If PC is between a NS and a subgiant,
it often leads to a direct LMXB formation
\citep{2005ApJ...621L.109I}.  In UCXBs, post-PC binary orbital
evolution is solely due to angular momentum loss via gravitational
wave radiation, same mechanism operates during the mass transfer, and
there is no known effects through which the metallicity could affect
the appearance of the X-ray binary during the mass transfer.  The
theoretically predicted number of UCXBs is consistent with the
observed bright LMXBs in MW~GCs, in case if at least several per cent
of ever formed NSs is retained \cite{2005ApJ...621L.109I}.

In extragalactic GC~LMXBs, which are generally more luminous than
``bright'' LMXBs in MW~GCs, the potential donors are likely RGs and
WDs.  For neither of the donors, as discussed above, metallicity
affects the formation rate {\it per giant}.  The dynamical properties
of blue and red extragalactic GCs in the studied sample were found to
be statistically identical \citep{2013ApJ...764...98K}.  The total
formation rate depends however also on the number of encounter
participants, which are NSs and RGs.  Theoretically, it is expected
that slightly more of NSs can be formed in a metal-poor GC than in a
similar metal-rich GC \citep{2008MNRAS.386..553I} -- this trend is
opposite to the observed metallicity dependence.

From stellar evolution, we know that metal-poor MS stars evolve faster
than metal-rich MS stars of the same mass.  Therefore, metal-poor GCs
of the same age have stars of a lower mass at their MS turn-off.
Lifetime of the RG stage is shorter for metal-poor stars as well.
Since RGs live shorter, a metal-poor GCs should contain RGs that have
a smaller mass range than RGs in a metal-rich GCs.  If one assumes
that GCs of different metallicities had the same initial mass
function, detailed calculations confirm that the fraction of RG stars
that a metal-poor GC can have is about twice less than a metal-rich GC
has \citep{2012ApJ...760L..24I}.  This directly affects the encounter
rates, and increases it in a metal-rich GC, compared to a metal-poor
GC, by a factor of two.

The mass of the RG also plays a role: first, the expulsion of a RG
envelope, which is more massive in a metal-rich RG, results in a
smaller post-PC separation, and in a direct LMXB formation in a larger
fraction of encounters; and second, the effective cross-section -- how
likely each RG is to participate in a dynamical encounter -- is higher
in metal-rich GCs.  In total, this results in about 3 times more
frequent formation of UCXBs in metal-rich clusters than in metal-poor,
and can explain the observed metallicity dependence
\citep{2012ApJ...760L..24I}.  The RG lifetime affects RG-LMXBS as
well, as the duration of a RG-NS mass transfer is shorter in
metal-poor GCs.

Between all GC LMXB classes, the best theoretical understanding, as
well as the best agreement on numbers with the observations of
MW~GC~LMXBs, we have is for the formation and evolution of UCXBs.  It
is thought that millisecond pulsars (MSPs) are related to LMXBs, and
transitional systems between LMXBs and binary MSPs (bMSPs) are now
known \citep[e.g.,][]{1998Natur.394..344W,2009Sci...324.1411A,2013Natur.501..517P}.
Simulations of GCs have shown good agreement between the number and
period distribution for all theoretically produced radio bMSPs, but
with the striking exception for radio bMSPs presumably produced by
UCXBs \citep{2008AIPC..983..442I}.  The lifetime of an UCXB at
$L_x>10^{36}$~erg~s$^{-1}$ is $\tau\sim10^8$ yr. This implies that for
each currently observed bright UCXB, 100 time more bMSPs should have
been created, or about 600~bMSPs with a small orbital period and a
low-mass companion should have been detected in MW~GCs.  However, we
do not observe such a radio bMSP population in GCs, nor in the overall number of MSPs,
nor for expected companion-mass and orbital period parameter space for bMPSs.
This problem is by no means confined only to GCs, there is a lack of
post-UCXBs radio bMSPs in the galactic ``field'' population as well
\citep{2008AIPC..983..501D}.

The interesting case that can provide in future a clue on the fate of
post-UCXB systems is the discovery of PSR J1719-1438 -- a ``field''
MSP which has the orbital period of 2.2 hour, its companion mass is
about that of Jupiter, but the minimum companion density is 23 g
cm$^{-3}$, suggesting that it may be an ultralow-mass carbon white
dwarf \citep{2011Sci...333.1717B}. This system is very puzzling, as it
takes more than a Hubble time to get to this period under conventional
UCXB evolution \citep{2012A&A...541A..22V}, while this evolution can
be deemed possible if additional donor wind, irradiation and donor's
evaporation are taken into account \citep{2012ApJ...753L..33B}.  While
this may explain why we do not see post-UCXB system as bMSPs, it still
does not explain the mismatch in the total number of MSPs.  On the other
hand, it is plausible that NSs that were spun up in an UCXB binary are less likely to be detected in radio,
and instead can be detected in $\gamma$-rays, as, for example, in case
of PSR~J1311-3430 that has a 93 min orbital period and very low-mass
helium companion \citep{2012Sci...338.1314P,2012ApJ...760L..36R}.

\section{Role of triples}

\label{triples}

In a hierarchical triple, a distant third body exerts tidal forces on
the inner binary. As a result, there is a cyclic exchange of the
angular momentum between an inner binary and a third body, causing
variations in the eccentricity and inclination of the stars orbits
\citep{1962AJ.....67..591K,2000ApJ...535..385F,2002ApJ...578..775B}.
If Kozai mechanism is coupled with dissipative tidal friction (Kozai
Cycle with Tidal Friction, KZTF), the inner binary can be driven to
start the RL overflow of one of the companions, with the subsequent
either a merger or a start of the stable mass transfer
\citep{2006Ap&SS.304...75E,2007ApJ...669.1298F}.  If the timescale of
KZTF is shorter than the timescale of dynamical encounters, KZTF
can provide a formation channel for LMXBs in GCs.

The interesting case is LMXB 4U~1820-303 in NGC~6624.  Binary orbital
period of this system is $\sim 685$s
\citep{1987ApJ...312L..17S,1997ApJ...482L..69A}.  Theoretically
predicted period increase for this system, if it is a binary, is $\dot
P/P> 8\times10^{-8}$ yr$^{-1}$ \citep{1987ApJ...322..842R}.  However,
from observations, the orbital period is decreasing as $(\dot
P/P)_{\rm obs}=-(5.3\pm0.3)\times 10^{-8}$yr$^{-1}$
\citep{1993A&A...279L..21V,2014ApJ...795..116P}.  In addition, 4U~1820-303 has the
luminosity variation by a factor of $\sim 2$ at a super-orbital period
$P\sim170$d \citep{2001ApJ...563..934C}.  \cite{2001ApJ...563..934C}
have suggested that this system could be a hierarchical triple with a
1.1 day outer orbital period.

\cite{2012ApJ...747....4P} have found that they can explain this
system to be a triple, where the inner binary is composed of
$1.4M_\odot$ NS and $0.067 M_\odot$ WD, and the third companion has
mass of $0.55 M_\odot$.  However, the outer semi-major axis is small,
$1.5 R_\odot$, and the outer orbital period is only 0.15d.  The
re-analysis of simulations performed in \cite{2008MNRAS.386..553I}
have shown that none of the dynamically formed triples with a NS was
anywhere close to the observed compactness, with the tightest
dynamically formed triple having its outer orbital period about a day,
and most of dynamically formed triples were even much wider.  This
should come as no surprise, as an inner binary in a dynamically formed
hierarchical triple would be rarely substantially tighter than the
tightest of the two binaries that have participated in the encounter.
Too tight binaries both have a small encounter rate and can be easily
destroyed via collisions during an encounter
\citep{2004MNRAS.352....1F}.  A possible explanation is that this
hierarchical triple, formed initially dynamically, was substantially
wider than now, and then was tightened through either {\it external}
common envelope event, initiated by an outer companion, or a {\it double}
common envelope event. The latter is a bit less likely, also in that
case the seed inner WD-NS binary could not have been formed via a PC.

\cite{2014arXiv1411.0368P} had also suggested that 3 more GC UCXBs
could be triple systems.  This, if confirmed, would bring the fraction
of triple UCXBs to 2/3 of all MW~GC~UCXBs.  It also may suggest that
yet unknown formation channel is as efficient as the current favorite
formation channel of UCXBs -- PC, or that most of NS-WD binary, formed
initially via a PC, form subsequently a triple system.  How plausible is that?
Theoretically expected formation rates suggests that each NS binary
has a 5\% chance to form a hierarchically stable triple in a dense GC
like 47~Tuc per Gyr, and as much as 15\% chance per Gyr in a cluster
like Ter~5. So, in a dense cluster, most of seed NS binaries could
become a member of a triple system at some point
(\citealt{2008msah.conf..101I}, also reanalysis of simulations
presented in \citealt{2008MNRAS.386..553I}).  The formation of the
triples with very short outer orbital periods is however as difficult
dynamically as for the described above 4U~1820-303.

\section{LMXBs with BH accretors}
 
Observational frequency of very bright extragalactic GC~LMXBs, with
$L_x \ga 10^{39}$ erg s$^{-1}$ indicating possible BH accretors, is
about $0.7\div 2\times10^{-9}$ per $M_\odot$ in massive GCs with an average $M_{\rm V}\approx -9$
\citep{2006ApJ...647..276K,2007ApJ...660.1246S,2008ApJ...689..983H,2013ApJ...764...98K}.
BH LMXBs formed via binary exchanges are likely to have low duty
cycles and not be detected as bright LMXBs
\citep{2004ApJ...601L.171K,2008ApJ...683..346F,2009ApJ...702L.143F}.
TCs by a BH likely destroy the capturing star
\citep{2004ApJ...601L.171K}. This leaves as most likely donors WDs.  A
BH-WD LMXB spends an order of $5\times10^5$ yr at $L_x \ga 10^{39}$
erg s$^{-1}$.  Assuming that about 10\% of all formed BHs is retained,
the observationally inferred formation rate of LMXBs is $\sim 4\times
10^{-3}$ per Gyr per each retained BH.

It has been shown that PCs with giants, the most effective mechanism
to form NS-WD LMXBs, can not provide direct formation of BH-WD LMXBs
at the observationally inferred rate.  A dynamically formed 
binary with a BH is several times more massive than any non-BH binary
in a GC.  If a dynamically formed BH-WD binary participates in a
binary-binary encounter, the chance to form a hierarchically stable
triple is substantial.  If Kozai timescale is short compared to the
characteristic time this triple has before its next dynamical
encounter, triple induced mass transfer can transform the inner BH-WD
binary into LMXBs \citep{2010ApJ...717..948I}.  Interestingly,
\cite{2010MNRAS.409L..84M} have found a large X-ray flux variation in
RZ~2109, such that this source is consistent with being in a triple.
RZ~2109 was the first discovered extragalactic GC X-ray source which
is believed to contain a stellar mass BH \citep{2008ApJ...683L.139Z}.

The observations of bright BH LMXBs in extragalactic GCs can be
extrapolated to our MW~GCs.  It is important that after the ``bright''
stage at $L_x \ga 10^{39}$ erg s$^{-1}$, an ultra-compact X-ray BH-WD binary
will remain a mass-transferring system for many
Gyrs.  With uniform formation rate, and average GC age of 10
Gyr, an average BH-WD LMXB will exist $\sim 10^4$
times longer than a bright BH-WD LMXB, and will spend most of its life
at extremely low mass transfer rates.  The expected frequency per
dense and massive MW~GCs is then of order $10^{-5}$ per $M_\odot$.
This implies that almost every massive GC in the Galaxy might contain
a very faint BH-WD LMXB! Note that only 7 MW~GCs are above $M_{\rm V}=-9$, 
to have BHs retention comparable to massive extragalactic GCs.

The recent radio observations of MW~GCs clusters have indeed indicated
that some of them contain faint potential BH-LMXBs.  Five candidates
are under investigation -- two LMXBs in M22
\citep{2012Natur.490...71S}, one in M62 \citep{2013ApJ...777...69C},
in 47 Tuc (Miller-Jones et al. in prep.)  and in one more GC
(Shishkovsky et al. in prep). Interesting that only in two cases WD
companions are plausible, in M22 and in 47~Tuc. The rest of the
companions are likely non-degenerate -- MS stars, or giants, or ``red
stragglers''. A BH LMXB with a non-degenerate donor can continue mass
transfer at a very low rate for several Gyrs, comparable to that of
BH-WD binaries, hinting that the dynamical formation rate of BH
binaries with non-degenerate donors is comparable to that of BH~UCXBs.
The formation booster with help of triples has not yet been shown to
work effectively for non-degenerate companions.  Therefore, the puzzle
of the formation of BH~LMXBs with non-degenerate companions in GCs
remains to be explained.



\bibliographystyle{mn2e}

\end{document}